\documentstyle[epsfig,referee]{mn}

\begin{document}

\title[Constraints on Star Formation]{Constraints on Star Formation from
the Close Packing of Protostars in Clusters}

\author[Bruce G. Elmegreen, and Mohsen Shadmehri]
{B. G. Elmegreen\thanks{E-mail: bge@watson.ibm.com} and Mohsen
Shadmehri\thanks{E-mail: mshadmehri@science1.um.ac.ir}\\ IBM
Research Division, T.J. Watson Research Center, P.O. Box 218,
Yorktown Heights, NY 10598, USA, and \\Department of Physics,
School of Science, Ferdowsi University, Mashhad, Iran}

\date{Received 20 August 2002 / Accepted  }

\maketitle

\markboth{Elmegreen \& Shadmehri a: Constraints on Star
Formation}{}

\begin{abstract} The mm-wave continuum sources (MCS) in Ophiuchus
have mutual collision rates less than their collapse rates by a
factor of 10 to 100, suggesting most will form stars without
further interactions.  However, the ratio of these rates would
have exceeded unity in the past if they were only $2.5$ times
larger than they are now.  Such a high previous ratio suggests
three possible scenarios: (1) the MCS contracted from lower
densities and acquired their present masses through collisional
agglomeration, (2) they contracted independently from lower
densities elsewhere and moved to the cluster core recently, or (3)
they grew from smaller sizes at a constant high density. The third
of these is most likely, implying that the MCS formed in the
shocked regions of a supersonically turbulent fluid.  The first
scenario gives the wrong mass function and the second does not
give the observed hierarchical clustering. The ratio of rates also
exceeds unity today if the MCS have envelopes with smooth profiles
that end in pressure balance with the ambient cloud cores; this
suggests again that turbulent flows define their outer
layers. Proximity constraints like this are even more important in
massive clusters, including globular clusters, in which massive
stars with the same or greater space density are more strongly
interacting than the Ophiuchus MCS.  As a result, the density
contrast for MCS must be larger in massive clusters than it is in
Ophiuchus or else significant
coalescence will occur in the protostellar phase, possible forming
massive black holes.  A proportionality to the second power of the Mach
number allows the MCS cores to collapse independently.  These
results suggest that stars in dense clusters generally form on a
dynamical time by the continuous collection and rapid collapse of
turbulence-shocked gas. Implications of proximity constraints on
the initial stellar mass function are also discussed. Warm cloud cores can
produce a top-heavy IMF because of a simultaneous increase in the thermal
Jeans mass and the collisional destruction rate of low mass MCS.
\end{abstract}


\section{Introduction}

The mm-wave continuum sources (hereafter MCS) in Ophiuchus (Motte,
Andr\'e \& Neri 1998; Johnstone et al. 2000), Serpens (Testi \&
Sargent 1998; Testi et al. 2000), and Orion (Johnstone et al.
2001) should reveal the process of their own formation. They
appear to be the last phase in the gas before it collapses into
stars. According to Motte et al., the typical density of an MCS in
the Oph B2 core is $\sim6\times10^7$ H$_2$ cm$^{-3}$, the size is
$\sim3000$ AU, the mass is $\sim0.3$ M$_\odot$, and the projected
separations are $\sim3$ arcmin $\sim0.1$ pc $\sim20,000$ AU.  This
separation corresponds to a spatial density on the order of $10^3$
pc$^{-3}$, considering that the overall core size is about the
same 0.1 pc. This density is comparable to that in embedded star
clusters (McCaughrean \& Stauffer 1994; Carpenter et al. 1997).

The relative velocities between the MCS are probably $\sim2^{1/2}$
times the three-dimensional virial speeds in their cloud cores,
which for Oph B2 is $\left(6GM/5R\right)^{1/2}\sim2$ km s$^{-1}$
for a core mass of $M=50$ M$_\odot$ and a core radius $R=0.055$ pc
(Motte et al.). Sources outside of dense cores may have smaller
velocity dispersions (Testi et al. 2000; Belloche, Andr\'e, \&
Motte 2001). If the dispersion is small and the MCS are far apart,
then they will not interact before they collapse into stars and
the final stellar mass function will be about the same as the MCS
mass function. However, if the dispersion and density of MCS in a
cloud core are high enough, then some may collide before they
collapse, and this can affect the mass function (e.g., Bonnell,
Bate, \& Zinnecker 1998).

The relative collision rate should have been higher in the past if
the MCS passed through a lower density state. The collision rate
scales approximately with the square of the MCS radius, $R^2$, and
the collapse rate scales with $R^{-3/2}$, from the square root of
density with mass conservation.  The ratio of the collision rate
to the collapse rate therefore scales with $R^{7/2}$, which is a
sensitive function of MCS size.  This implies that even if the MCS
are not collision-dominated now, they did not have to be much
larger in the past before their internal dynamical evolution was
severely affected by interactions.  If such interactions were
destructive for the MCS or detrimental to their stellar-like mass
function (Motte et al. 1998), then they could not have formed by
contraction from the cloud core through a sequence of nearly
constant mass, but had to form from smaller masses by coagulation or
accumulation through a sequence of near-constant density,
or they had to form elsewhere.

Here we determine the relative collision rates for the MCS in core
regions B and C of Ophiuchus.  We also determine the relative
rates for idealized MCS with radial density profiles from
Whitworth \& Ward-Thompson (2001) and a mass distribution from the
initial stellar mass function (IMF).

The ratio of the collision rate to the collapse rate increases
with the mass of an MCS.  This means that massive clusters with
the same or higher densities will have more severe crowding
problems than the cores in Ophiuchus, which is making only
low-mass stars.  This implies that the contrast between the gas
density in an MCS and the gas density in the ambient cloud core
must increase with cluster mass or density.  This effect is
modelled in Section \ref{sect:highmass}.

An important simplification in this problem is that collisional
destruction of MCS should be about the same for interactions with
other MCS as it is for interactions with protostars and stars.
This is because the grazing collision distance between one MCS of
mass $M_i$ and another of mass $M_j$ having the same density is
equal to the tidal distance for the destruction of the first MCS
by a point source of mass $M_j$. Tidal destruction occurs when an
MCS enters a region where the average density is the same as its
own density or higher. Thus our result is unchanged if only a
fraction of the final stars are present as MCS at any one time.
Most likely, there is a continuous conversion of existing MCS into
stars while new MCS are forming out of the ambient cloud core gas.
The ratio of the numbers of MCS to final stars will be about the
same as the ratio of the dynamical times, which is the inverse
square root of the density contrast.

We also consider whether the turnover mass in the IMF might depend
on the cluster core density, taking higher values for denser
clusters because of limitations from crowding. The effect is
probably small, but it may be observable under extreme conditions.
For example, coalescence in the protostellar phase of very dense
clusters could shift a significant fraction of low-mass MCS to
higher mass. Considering how close the Salpeter IMF is to an
$M^{-2}dM$ form (this form has equal total mass in equal
logarithmic intervals of mass), this shift means that an IMF which
extends as a Salpeter power law to low mass but falls below the
Salpeter power law at high mass in a low density environment might
shift to one that turns over at low
mass and has a Salpeter power-law slope at high mass
in a high density environment. Observations
of the IMF support these changes with stellar density, although
other explanations are possible.

\section{The Relative Collision Rate}
The collision rate of an MCS is determined from the space density
of its neighbors, the relative velocity dispersion, and the
gravitationally-enhanced collision cross section. The collapse
rate is determined from the MCS density. The ratio of these two
rates indicates the relative probability of collisions versus
isolated collapse. If the ratio exceeds unity then collisions
should be important. Here we use the observed sizes and masses for
MCS tabulated by Motte, Andre \& Neri (1998) and we determine the
ratio of these rates to see how important collisions are. We then
consider the same ratios again for the MCS with hypothetically
lower densities.
The results are then generalized to the case in which the number
of MCS is given by the initial mass function and each MCS has an
internal density profile from Whitworth \& Ward-Thompson (2001).

\subsection{Observed mm-wave continuum sources in Ophiuchus}

A field of $N$ MCS with radii $R_j$, mass $M_j$, and uniform
velocity dispersion $v$ inside a cloud core with radius
$R_{cloud}$ and mass $M_{cloud}$ has grazing (hard-sphere)
collisions with a particular MCS of radius $R_i$, mass $M_i$, and
velocity $v$ at the rate
\begin{equation}
\\ \omega_i={\pi\over {4\pi R_{cloud}^3/3}} \sum_{j=1}^{j=N}
\left(R_i+R_j\right)^2 2^{1/2} v \left[1+
{{2G\left(M_i+M_j\right)}\over{2v^2\left(R_i+R_j\right)}}\right].
\label{eq:omegai}
\end{equation}
The virial speed in the cloud core gives the three-dimensional
velocity dispersion,
\begin{equation}
\\v=\left(3GM/5R\right)^{1/2},
\end{equation}
which is an approximation valid for a uniform cloud; the extra
factors of $2^{1/2}$ near $v$ and $2$ near $v^2$ in equation
(\ref{eq:omegai}) are from relative motion assuming the same speed
for all MCS. The collapse rate for a uniform MCS of density
$\rho_i=3M_i/\left(4\pi R_i^3\right)$ is
\begin{equation}
\\ \omega_{\rm collapse,i}=\left(32G \rho_i /3\pi\right)^{1/2}.
\end{equation}

The ratios of the collision rates to the collapse rates for MCS
in the B and C cores of Ophiuchus are shown in Figure 1, plotted as
solid symbols versus the MCS mass.  The ratios are in the range
from 0.001 to 0.1, increasing slightly with mass for Oph B. Such
low values indicate that collisions are slower than internal
dynamical processes. If the MCS are gravitationally unstable to
collapse into stars, then they will do this independently, without
interactions.

\begin{figure}
\epsfig{figure=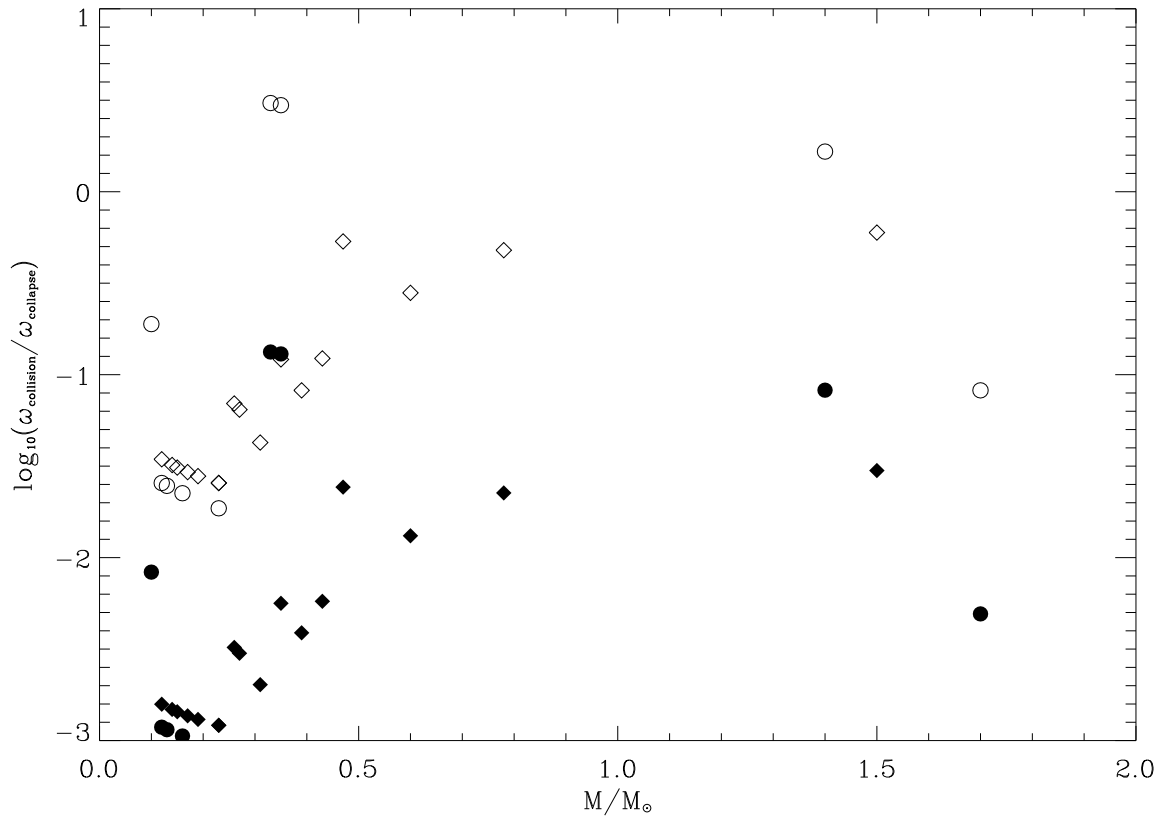,angle=0,width=\hsize} \caption{The
ratios of collision rate to collapse rate for mm-wave continuum
sources (MCS) in Ophiuchus are shown versus the MCS masses using solid
symbols.  The source Oph B2 is indicated by diamonds and Oph C by circular
symbols.
The ratios increase slightly with mass but are generally
less than one.  The ratios are shown again as open symbols for MCS
radii larger than observed by a factor of 2.5. Even this small
change in radius would cause several of the MCS to collide or
coalesce before they could collapse into a star, as the
plotted ratios become larger than 1.
This suggests that the MCS did not pass through a lower density
stage with their present masses.} \label{fig1}
\end{figure}

The same ratios are plotted again in Figure 1 as open symbols for
a hypothetical previous state in which each MCS was larger than it
is today by a factor of 2.5.  The MCS masses are assumed to be
unchanged.  A small difference in MCS size corresponds to a large
difference in relative collision rate, so with larger sizes, some
MCS have ratios of rates larger than 1.  These MCS could not have
reached their present state by purely dynamical collapse without
strongly interacting with previous neighbors. Internal evolution
that is slower than the collapse rate makes previous interactions
even more likely, as would the presence of a larger number of MCS
in the previous cloud core.

The dynamical time scale in the present-day MCS is only $\sim4000$
years at the average MCS density of $\sim6\times10^7$ H$_2$
cm$^{-3}$. The dynamical time in each whole core is $\sim10$ times
longer because of the $100\times$ lower average core density. If
the MCS and the cores around them both evolved dynamically on
their own time scales, then the MCS would have had twice their
present radii only several $\times10^4$ years ago, when the cores
as a whole were not much different than they are today.
This is such a short time ago compared to the age of the whole
Ophiuchus star forming region that it seems unlikely the
MCS recently suffered some collision without producing
tidal features such as bridges or tails. Most likely {\it the
MCS have not been in a lower density state since they were
part of the ambient cloud core gas}.

\subsection{Protostars with Smooth Density Profiles}

The MCS in Ophiuchus are denser than the average cloud cores by a
factor of $\sim100$, suggesting a mismatch in thermal pressure.
The boundaries of the MCS must therefore have either a gradual
density gradient, taking the MCS density down to the cloud core
density, or a shock front at which the core thermal pressure
balances an inter-MCS ram pressure from relative motions or
turbulent flows. Boundaries
like this are required whether or not the MCS are
self-gravitating.

Dense mm-wave sources are often observed to have density gradients
like the one fitted by Whitworth \& Ward-Thompson (2001), which is
\begin{equation}
\\ \rho(r)={{\rho_{\rm flat}}\over{\left[1+\left(r/R_{\rm
flat}\right)^2\right]^2}}, \label{eq:rhor}
\end{equation}
for radius $R_{\rm flat}$ and density $\rho_{\rm flat}$ in the
central region where the profile flattens. The density contrast is
represented by
\begin{equation}
\\ {\cal C}=\rho_{\rm flat}/\rho_{\rm s}, \label{eq:eps}
\end{equation}
where $\rho_{\rm s}$ is the surface density of each MCS. The
parameter ${\cal C}$ depends on the definition of the surface. We
assume that the MCS surface occurs where its density equals the
average density in the cloud core (e.g., ${\cal C}=50$ is the
ratio of the average MCS density to the average cloud core density
in Oph B2; ${\cal C}=187$ is the average for Oph C -- Motte et al.
1998).

Equations (\ref{eq:rhor}) and (\ref{eq:eps}) give the radius of
the MCS, $R_s$, as a function of ${\cal C}$ and $R_{\rm flat}$,
\begin{equation}
\\R_{\rm s}=a R_{\rm flat} \;\;{\rm for}\;\;
a=\left({\cal C}^{1/2}-1\right)^{1/2}.\label{eq:rs}
\end{equation}
For mass
\begin{equation}
\\M=\int_{0}^{R_{\rm s}}4\pi r^{2}\rho(r) dr,
\end{equation}
\begin{equation}
\\R_{\rm flat}=\left(\frac{M}{2\pi\rho_{\rm
flat}}\right)^{1/3}\left(\arctan(a)-{{a}\over{1+a^2}}\right)^{-1/3}.
\label{eq:rf}\end{equation}

The average density of an MCS is
\begin{equation}
\\ \rho_{\rm av}=\frac{3M}{4\pi R_{\rm s}^3}=\frac{3\rho_{\rm flat}
}{2a^3}\left(\arctan(a)-\frac{a}{1+a^2}\right),
\end{equation}
giving an average collapse rate
\begin{equation}
\\ \omega_{\rm collapse,a}=\left(32G\rho_{\rm av}/3\pi\right)^{1/2}.
\end{equation}
The collapse rate in the flat part of the density profile is
\begin{equation}
\\ \omega_{\rm collapse,f}=\left(32G \rho_{\rm flat}
/3\pi\right)^{1/2}. \label{eq:omegaflat}
\end{equation}

A cluster of MCS with smooth density profiles has a higher
relative collision rate than what we have just calculated for the
MCS in Ophiuchus because smoothly tapered MCS are bigger and their
average densities are comparable to the cloud core density, giving
them a longer collapse time. For a general case, the collision rate for
an MCS depends on the space density of other MCS, which is a function of
mass through the IMF.

A convenient form for the IMF at low to moderate mass is
\begin{equation}
\\n(M)dM=n_{0}M^{-2.3}\left[1-\exp\left(-M^2/M_{\rm
t}^2\right)\right]dM ,\label{eq:imf}
\end{equation}
which has a Salpeter slope of $-2.3$ toward higher mass and a
turn-over toward lower masses at $M_{\rm t}$ (see Elmegreen 1997).
The turnover mass may equal the thermal Jeans mass unless crowding
limitations are important; we discuss these limitations in Section
\ref{sect:imf}.  We introduce a high-mass modification to the IMF
in Section \ref{sect:highmass} (see Fig. \ref{fig3}).

To get $n_{\rm 0}$, we set the lower mass limit from observations
of Ophiuchus, we equate the total MCS mass in the IMF to the
observed total, and we constrain the number of MCS with a mass
greater than some maximum mass to be unity.  In this way, the IMF
has the correct total mass and it also falls off gradually at high
mass to give only one most massive star. The minimum mass, $M_{\rm
min}$ is set equal to 0.1 M$_\odot$ for both Oph B2 and Oph C (see
Fig. 1). The
total MCS masses in both Ophiuchus B and C are equal to the
fraction $\epsilon=0.13$ times the cloud core masses.

\begin{figure}
\epsfig{figure=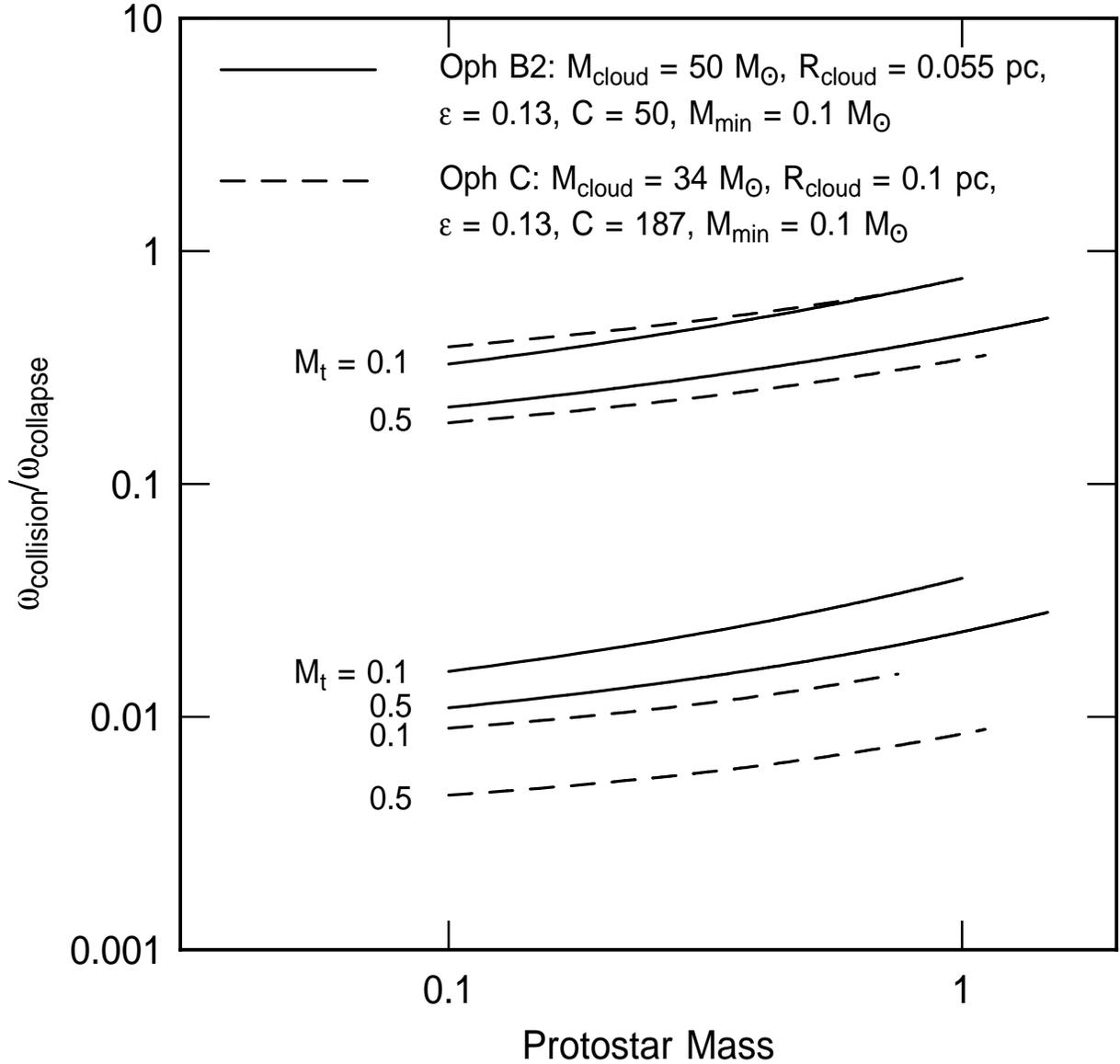,angle=0,width=\hsize} \caption{The
ratio of collision to collapse rates is shown versus the MCS mass
for an idealized case representative of the cores in Ophiuchus.
The MCS masses are distributed according to the IMF and they have
smooth density profiles. The lower four curves are for the flat parts
of the MCS only, while the upper four curves include the envelopes
also. $M_t$ is the turnover mass (in $M_\odot$) in the IMF. } \label{fig2}
\end{figure}

The collision rate for MCS $i$ is now obtained by integrating over
the IMF,
\begin{equation}
\\ \omega_{\rm collision} \left(M_i\right)={\pi 2^{1/2} v\over {4\pi
R_{cloud}^3/3}} \int_{M_{min}}^{\infty}
n(M_j)\left(R_i+R_j\right)^2
\left[1+{{2G\left(M_i+M_j\right)}\over{2v^2\left(R_i+R_j\right
)]}}. \right]dM_j \label{eq:colrate}
\end{equation}

The results of this model are shown in Figure 2.  The ratios of
the collision rates to the average collapse rates for MCS are
plotted as a function of MCS mass.  Solid lines are for an
idealized model of Oph B and
dashed lines are for Oph C.  Two different values of the turn-over
mass in the IMF, $M_t$, are assumed. The lower set of curves is
for
the MCS cores only, which come from
the flat part of the Whitworth \& Ward-Thompson profile only.
That is, it uses $\rho_{\rm flat}$ for the collapse rate, and
$R_{\rm flat}(M)$  for the collision cross section.

Figure 2 reproduces the ratios of the collision rates to the
collapse rates found in Figure 1 when only the flat parts of the
density profiles are considered.  These ratios are small, so the
flat parts should not collide with each other before they collapse
into stars. However, if the MCS have envelopes where the density
drops down to the average value in the cloud core, which
corresponds to an extra factor of $a\sim3$ in radius, then the
envelopes will interact significantly before they get involved
with the collapse. Thus {\it MCS in dense cluster environments should
not have significant envelopes}.

\section{Globular Clusters}
\label{sect:highmass}

Other models were considered for more massive clusters. When the
cluster mass is large, the largest MCS mass is also large, forcing
us to modify the IMF so that, for example,
1000 M$_\odot$ stars do not form. We
follow the discussion in Elmegreen (2000) and use for this case an
IMF of the form:
\begin{equation}
\\n(M)dM=n_{0}M^{-2.3}\left[1-\exp\left(-M^2/M_{\rm
t}^2\right)\right]\exp\left(-M/100 \; {\rm M}_\odot\right) dM
,\label{eq:imf2}
\end{equation}
which has a cutoff at high mass. The IMFs in equations
(\ref{eq:imf}) and (\ref{eq:imf2}) are shown in Figure 3
(multiplied by $M$ to convert to a $\log M$ coordinate). The
dotted line is from equation (\ref{eq:imf}) for $M_t=0.1$; it
blends with the solid line at low mass, which is from equation
(\ref{eq:imf2}) using the same $M_t$. The dashed line is from
equation (\ref{eq:imf2}) using $M_t=1$ M$_\odot$ and the same
total cluster mass as the solid line. The two straight lines show
slopes of $-1.3$ and $-1.6$ for comparison; these bracket the IMF
observations in the indicated mass range. There are no
observations of an IMF for masses much larger than 100 M$_\odot$,
so the fall off is purely conjectural at this time.

\begin{figure}
\epsfig{figure=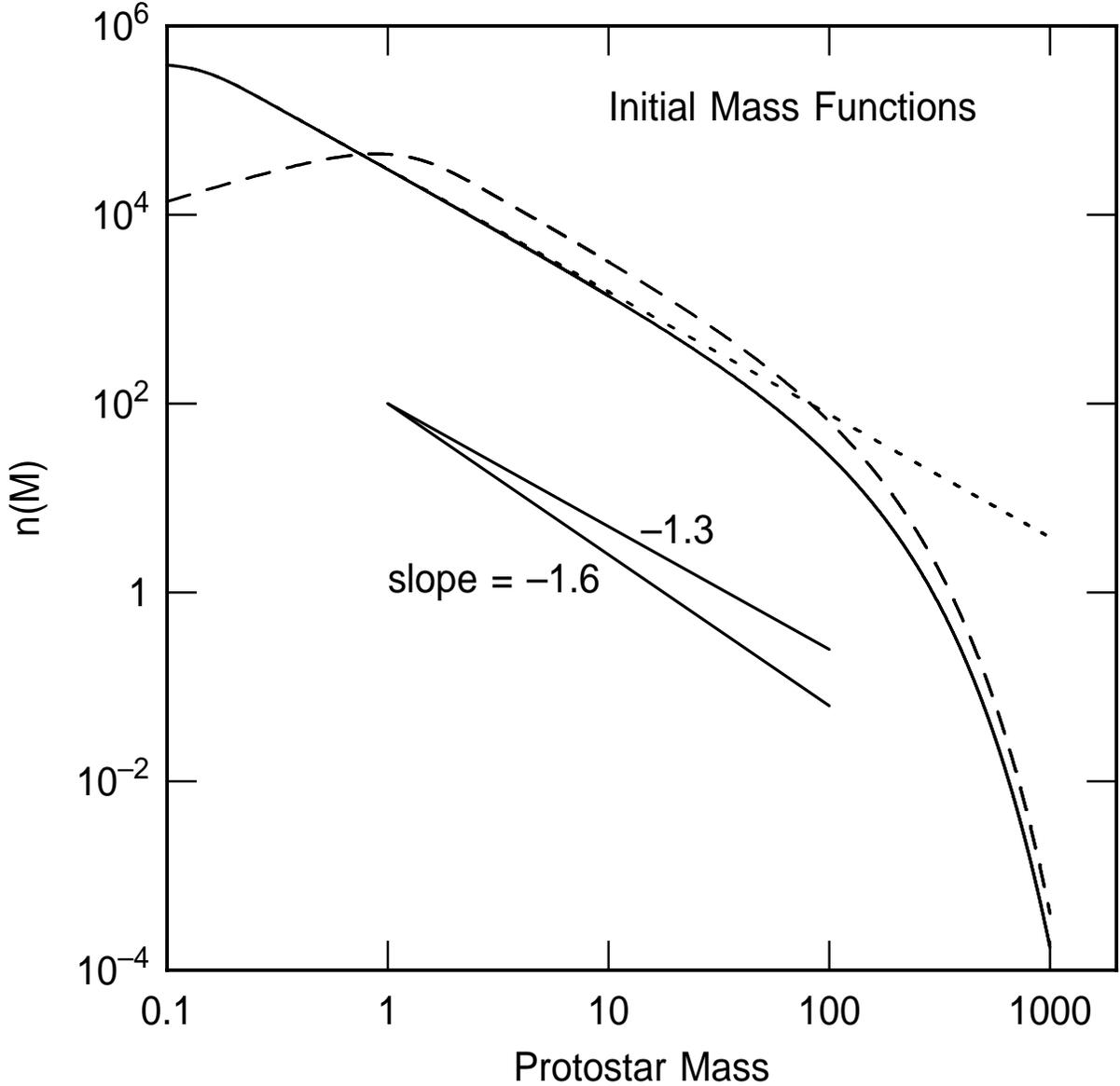,angle=0,width=\hsize} \caption{Models
for the initial stellar mass function used in the figures here.
The solid line has a turnover mass $M_t=0.1$ M$_\odot$ and the
dashed line has $M_t=1$ M$_\odot$.  Both have upper mass turnovers
too, while the dotted line is a power law to arbitrarily high
mass. } \label{fig3}
\end{figure}

The result for massive clusters are shown in Figures 4 and 5.
Figure 4 shows the ratio of rates versus MCS mass for two clouds
with $M_{\rm cloud}=6000$ and $6\times10^5$ M$_\odot$, and with
equal star formation efficiencies ($\epsilon=0.3$) and densities
($\rho_{\rm cloud}=7.8\times10^{-20}$ gm cm$^{-3}$), corresponding
to radii of $R_{\rm cloud}=$ 1.08 pc and 5 pc, respectively.  The
density contrast, ${\cal C}=160$, was also taken to be the same in
each case, and comparable to that in Ophiuchus. Two values of the
turnover mass, $M_t$ are considered, while $M_{\rm min}= 0.1$
M$_\odot$ for all cases.

\begin{figure}
\epsfig{figure=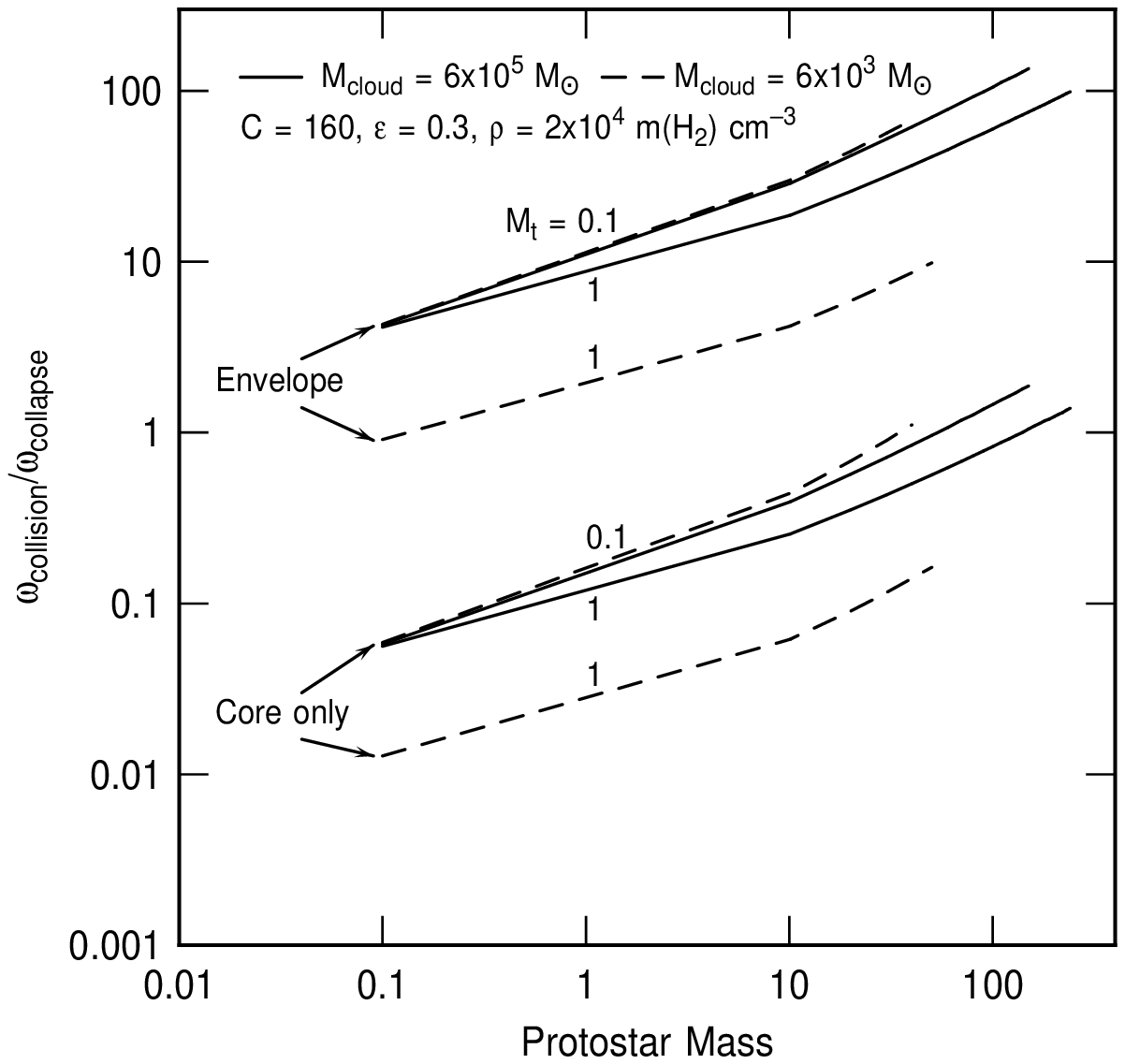,angle=0,width=\hsize} \caption{The
ratio of rates for two clusters more massive than Ophiuchus,
showing an increase in the ratios to unacceptably high values for
$\sim10^5$ M$_\odot$ clusters, even for the cores (i.e., the flat
parts of the density profiles), when the density contract ${\cal
C}$ is as low as observed in Ophiuchus.} \label{fig4}
\end{figure}

Figure 4 shows several things. First, the maximum mass increases
with the total stellar mass, $\epsilon M_{\rm cloud}$, as expected
for a statistical ensemble (i.e., the solid lines go to larger
protostellar mass than the dashed lines). Second, the ratio of
collision to collapse rates increases with MCS mass because of the
increasing collision cross section. Third, the ratio is much lower
for the MCS cores than for envelopes, as in Figure 2; it is also
lower for higher turnover mass, $M_t$. The reason for the $M_t$
dependence is that higher $M_t$ corresponds to relatively fewer low
mass stars, and this decreases the number density for collisions
in the field.

Figure 4 suggests a problem with star formation in massive
clusters. If the densities of the MCS-equivalent objects are the
same as in the Ophiuchus cores, then the envelopes of the massive
stars will strongly interact with most of the lower mass MCS,
probably accreting them.  Even the cores of the MCS in the highest
mass clouds (solid lines in the figure) will interact (by tidal
forces) strongly with other MCS and with the protostars already
formed. This implies that {\it massive clusters cannot have MCS with
the same densities as those in Ophiuchus} (i.e., $10^7$ -- $10^8$
m(H$_2$) cm$^{-3}$).

\begin{figure}
\epsfig{figure=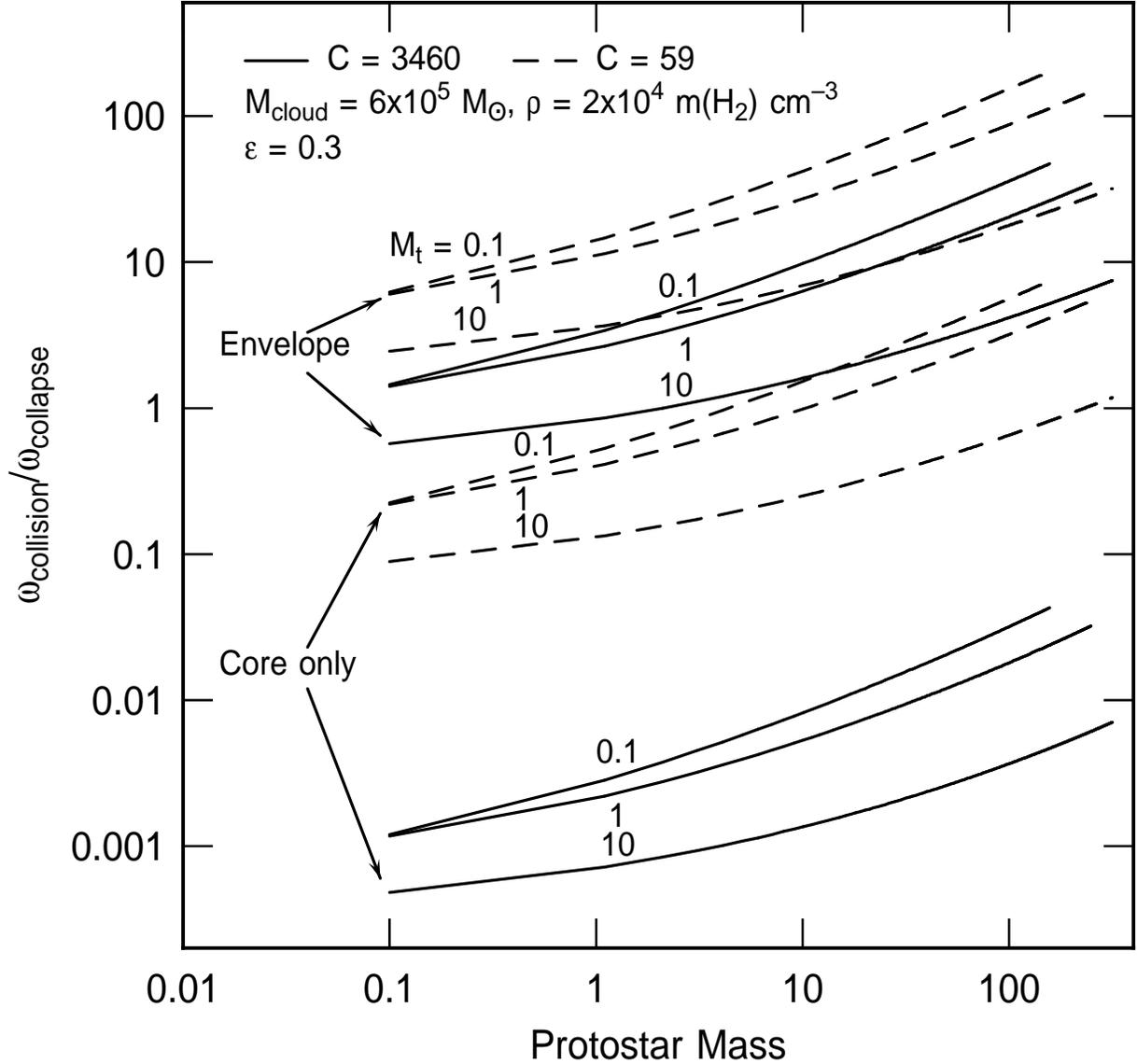,angle=0,width=\hsize} \caption{The
ratio of rates for a $2\times10^5$ M$_\odot$ cluster, like a
globular cluster, with two cases for the density contrast factor.
Only when the density contrast factor is very large, scaling with
the square of the cloud Mach number in this model, can MCS and
protostars co-exist in a cluster environment without destructive
interactions. } \label{fig5}
\end{figure}

Figure 5 shows the same clouds and clusters as in Figure 4 but now
with compression factors (${\cal C}= 59$,  3460) that scale with
the first (dashed lines) and second (solid lines) powers of the
Mach number from virialized motions, assuming a sound speed of 0.3
km s$^{-1}$. The assumption here is that MCS might be produced in
self-gravitating shocks that have average compression factors
proportional to some power of Mach number.  This power is likely
to be close to 1 when the shock moves across magnetic field lines,
and 2 when the shock moves parallel to the field lines.  The
shocks presumably arise from supersonic turbulence in the cloud
cores where MCS form, so the turbulent speed is taken to be the
cloud virial speed, which is the same as $v$ in equation
(\ref{eq:colrate}).

Figure 5 also shows cases with very large turnover masses,
$M_t=10$ M$_\odot$. When $M_t\le1$ M$_\odot$, {\it only the
Mach-squared scaling is strong enough to make the MCS so small
that their cores avoid severe collisions}. In all cases, the
envelopes collide. McKee \& Tan (2002) discuss
the last stages of massive star formation and note that it
requires very high initial densities and pressures. This is
consistent with the high value of ${\cal C}$ found here from
close-packing constraints in the massive clusters where massive
stars tend to form. One implication of this result is that {\it the
formation of stars from MCS in massive clusters should be faster
than in low mass clusters}.

\section{IMF Dependence on Cluster Mass and Density}
\label{sect:imf}

Figures 4 and 5 suggest that the MCS mass where collisions are
important ($\omega_{\rm collision}/ \omega_{\rm collapse} \sim 1$)
becomes insensitive to the IMF turnover mass, $M_t$, when the
ratio between the maximum stellar mass and $M_t$ is large. Once
this ratio exceeds several hundred, most collisions are
between MCS with masses above $M_t$ and then the value of $M_t$
does not matter.

Higher $M_t$ means fewer low-mass collision partners and more
freedom for the massive MCS to collapse without interference from
neighbors. For a given modest MCS density contrast, ${\cal C}$,
denser and more massive stellar clusters should have higher IMF
turnover masses because of crowding limitations.  However,
sufficiently large ${\cal C}$ gives all cores a low ratio of
collision to collapse rates, and then the IMF does not depend on
crowding.

An increase in thermal temperature in a cloud core should decrease
the density contrast in shocks and decrease ${\cal C}$.  This
increases the relative importance of collisions and can increase
$M_t$ by removing the low mass MCS.  The dependence of $M_t$ on
${\cal C}$ for a constant ratio of rates
is approximately $M_t\propto{\cal C}^{-2}$ from Figure 5. 
This comes from the mutual dependencies of 
$\omega_{\rm collision}/ \omega_{\rm collapse}$
on ${\cal C}^{-3.5/3}$ (see Sect. \ref{sect:scaling}) 
and $M_t^{-0.6}$ for the ``core only'' case with 
$M_t$ in the range from 1 to 10.  From these mutual
dependencies, 
$\omega_{\rm collision}/ \omega_{\rm collapse}$
is constant if $M_t^{-0.6}\propto {\cal C}^{3.5/3}$,
which gives $M_t\propto {\cal C}^{-1.94}$.

Such a ${\cal C}$ dependence is comparable to the dependence of the thermal
Jeans mass ($M_J$) on compression ratio, which would predict
$M_J\propto{\cal C}^{-2}$ for ${\cal C}$ proportional to the
square of the Mach number.
This comes from the dependence of the Jeans mass on 
$T^2$ for constant pressure, and the dependence of the
Mach number on $T^{-1/2}$ for constant virial speed in the cloud. 

The similar dependencies of $M_t$ and $M_J$ on ${\cal C}$
imply that the thermal Jeans mass
and the crowding limitations both
determine the turnover mass in the IMF.  If two
regions have different temperatures with all else the same, then
$M_t$ should be larger in the warmer region in proportion to the
thermal Jeans mass and the crowding conditions will
stay about the same.  This means that {\it warm cloud cores can produce
a top-heavy IMF because of a combination of increased thermal
Jeans mass and increased crowding}.

No clear dependencies between cluster density and IMF have been
found (Massey \& Hunter 1998), although some IMF observations
suggest they could be present. For example, the clustered embedded
stars in Orion have a shallower IMF than the unclustered embedded
stars (Ali \& DePoy 1995), and the LMC clusters in regions of high
stellar density tend to have shallower slopes than those in
regions of low stellar density (J.K. Hill et al. 1994; R.S.  Hill
et al. 1995). Similarly, the extreme field stars in the LMC have a
steeper IMF slope (Massey et al.  1995) than the dense and massive
cluster 30 Dor, which has a shallow slope like the Salpeter value
(Massey \& Hunter 1998). Shallow slopes correspond to
proportionally more massive stars, and this can be the result of a
previous stage in the evolution of the cluster where the low mass
MCS coalesced into higher mass MCS. The problem with these
interpretations is that other processes, such as mass segregation,
can change the slope of the IMF in the same way as MCS
coalescence.

Variations in $M_t$ have also been difficult to establish
observationally. The old globular clusters in the Milky Way halo
tend to have normal $M_t$, namely about half a solar mass (Paresce
\& De Marchi, 2000), whereas a young globular cluster in M82 has
$M_t$ higher than normal by a factor of $\sim 5$ (Smith \& Gallagher
2001).  The cluster masses and densities are about the same.

This lack of a clear correlation between the IMF and the density
or mass of a cluster (over and above possible changes in the
thermal Jean mass) implies that crowding effects are not
particularly
important in either the protostellar phase or in an earlier phase,
such as that producing the MCS in Ophiuchus.  This can only be the
case if the MCS that eventually form stars build up from smaller
objects at nearly constant density, and if this density scales
with at least the square of the Mach number in the cloud core.

Crowding and coalescence in the MCS phase will be very important in
the cores of globular clusters if the temperature is large or the
compression factor scales directly with the Mach number, as for a
strongly magnetic gas.  In this case, supermassive stars can form by
the coalescence of MCS. Such stars may either form massive black holes
directly, or they may continue to coalesce as stellar remnants until
several together make a massive black hole.  This is the
scenario for black hole formation in dense clusters
proposed by Ebisuzaki et al. (2001),
but modified slightly here to include the MCS phase, when collisions
are more important than in the main-sequence or post-main sequence phases.

\section{Scaling of the Results}
\label{sect:scaling}

The curves in Figures 4 and 5 have a similar dependence on MCS
mass that comes mostly from the mass dependence for the square of
the MCS radius, which is approximately as $M^{2/3}$.  Lower $M_t$
clusters have slightly higher ratios of rates because of the
correspondingly larger MCS spatial densities.  Lower $M_t$
clusters also produce slightly steeper slopes in the figures.  The
steeper slope at low $M_t$ arises because the terms $(R_i+R_j)$
and $(M_i+M_j)$ are dominated by the radius and mass of the test
object, $R_i$ and $M_i$, when most of the field MCS are at much
smaller radii and masses. This occurs at low $M_t$, considering
that the integral in equation (\ref{eq:colrate}) is dominated by
the field MCS mass at the peak of the IMF, which is close to
$M_t$.  When $M_t$ is high, most of the field interactors are
large and massive, and then the dependencies of $(R_i+R_j)$ and
$(M_i+M_j)$ on $R_i$ and $M_i$ alone diminish.  The test mass,
$M_i$, is the quantity plotted on the abscissa in the figures.

The vertical shifts between curves may be understood in terms of
the dependencies of the ratio of rates on the compression factor,
${\cal C}$, and the cloud mass, $M_{cloud}$.  In Figure 4, the
cloud mass in two cases differs by a factor of 100 while the cloud
density and the MCS densities and radii are the same.  The cloud
mass enters into the velocity dispersion $v$, the cloud radius,
$R_{cloud}$, and the total number of MCS, 
which appear in equation (\ref{eq:colrate}). 
For the assumed constant cloud density, the only remaining
dependence is on $v$, which scales as
$M_{cloud}^{1/3}$.  Thus the dashed and solid
lines for both core and envelope models are displaced vertically
in the figure by $100^{1/3}=4.64$, which is 0.66 in the log, as
observed.

In Figure 5, the compression factor in two cases differs by a
factor of 58.6 for the same cloud mass and radius.   For the
envelopes, the MCS radii scale with ${\cal C}^{1/4}$ times the
core radii, and the core radii scale with ${\cal C}^{-1/3}$ for a
given MCS mass.  Thus the MCS envelope radii scale with ${\cal
C}^{-1/12}$ at constant mass when ${\cal C}$ is large.  The ratio
of collision to collapse rates scales with radius as $R^{3.5}$,
which is as ${\cal C}^{-3.5/12}$.  Considering the ratio 58.6 for
${\cal C}$, this means that the dashed and solid curves for the
envelope case in Figure 5 should be vertically displaced by
$58.6^{-3.5/12}=0.30$, which is $-0.5$ in the log, as observed
approximately.  For the cores, the MCS radii scale with ${\cal
C}^{-1/3}$ for a given MCS mass and so the ratio of rates scales
with ${\cal C}^{-3.5/3}$.  This gives a vertical displacement
between the dashed and solid lines for the cores that is equal to
a factor of $58.6^{-3.5/3}=0.0085$, which is $-2.1$ in the log, as
observed.

The displacement between the envelope and core cases results
mostly from the ${\cal C}^{1/4}$ ratio between envelope and core
radii, given the $R^{3.5}$ dependence for the ratio of rates at
constant MCS mass. Thus in Figure 5, the vertical displacement for
the ${\cal C}=59$ case is $59^{3.5/4}=35.4$, which is $1.55$ in
the log, and the vertical displacement for the ${\cal C}=3460$
case is $3460^{-3.5/4}=1249$, which is $3.1$ in the
log.  Both are approximately correct compared to the dislacements
in the Figure.

\section{Conclusions}

The mm-wave continuum sources in Ophiuchus are not interacting
strongly at the present time and should collapse to stars
independently, preserving their masses (Motte et al. 1998).
Previous versions of these MCS should have interacted more
strongly if their densities were lower at fixed masses, and MCS in
more massive clusters should interact more strongly too.  If such
interactions are destructive, or if they modify the mass function,
then such a previous stage could not have occurred. This might
imply that the MCS formed elsewhere and fell into the cloud core
after they had their high densities.  However, such a model would
not explain the hierarchical positions of these MCS, as seen in
both Ophiuchus and Serpens (Testi, et al. 2000); MCS ballistic
motions over one or more core dynamical times would mix them up.

A high ratio of rates might also imply that collisions were
important in the past, particularly among the most massive MCS.
However, pure collisional agglomeration leads to a shallow mass
function, $M^{-1.5}dM$ (Field \& Saslaw 1965), not the observed
$M^{-2.3}dM$ function at high MCS mass (Motte et al. 1998; Testi
\& Sargent 1998; Johnstone et al. 2001). 
Accretion of ambient gas by moving protostars
may give a better mass function (Bonnell et al. 1998), but such
models are not collision-dominated like the high $\omega_{\rm
collision}/\omega_{\rm collapse}$ cases here.

{\it The most likely scenario for the formation of the MCS in Ophiuchus
and elsewhere is that they grow inside high-density turbulent
shocks and become unstable when they reach a sufficiently large
mass}.  Such a process would not have a previous stage where low
density MCS were strongly interacting.  It would also account for
the hierarchical positioning of MCS in cloud cores (because
turbulence makes hierarchical structure), the inferred Mach number
dependence of the MCS density with increasing cloud core mass, the
lack of smoothly varying envelopes, and the required rapid
formation time of the MCS, considering their high densities.
Models of turbulence-induced star formation are in
Elmegreen (1993), Padoan (1995), Heitsch, Mac Low, \& Klessen (2001),
Klessen, Heitsch, \& Mac Low (2000), Klessen (2001),
Ossenkopf, Klessen, \& Heitsch (2001), Padoan, et al. (2001), and
Wuchterl \& Klessen (2001).
Other formation scenarios might be preferred outside of dense clusters.

For turbulence-induced star formation, the number of stars that
finally forms in a cluster should be larger than the number of MCS
that are present at any one time because the MCS phase is shorter
than the lifetime of the whole core. The ratio of the number of
MCS to the number of final star counts should be the ratio of
corresponding evolution times, which is the inverse square root of
the ratio of densities, ${\cal C}^{-1/2}$. This ratio equals
$\sim0.1$ for the Ophiuchus cores and could be as small as $0.01$
for super star clusters ($M\sim10^5$ M$_\odot$). The
MCS-equivalent phase in massive clusters should also be associated
with higher densities, possibly in proportional to the square of
the Mach number.  Thus objects with densities of $10^7$ cm$^{-3}$
in Ophiuchus would appear in a similar phase of evolution having
densities of $\sim10^{9}-10^{10}$ cm$^{-3}$ in super star clusters.  Of
course, observations with a sensitivity to a particular density
will always see the observable value, so comparisons between low
and high mass clusters should look for the same pre-stellar phases
in terms of structure and kinematics
when comparing densities.

The IMF should be affected by crowding in the sense that warmer,
more crowded
environments should have a higher low-mass limit to the power law part
and proportionally more massive stars to conserve total mass. The
temperature dependencies of the thermal Jeans mass and the low mass
limit from crowding are about the same.

\end{document}